\documentclass[twocolumn,showpacs,preprintnumbers,amsmath,amssymb,prb,superscriptaddress]{revtex4-1}
\usepackage{mathrsfs}
\usepackage{graphicx}
\usepackage{dcolumn}
\usepackage{bm}
\usepackage{amsmath}
\usepackage{amsfonts}

\begin{document}

\title{Calculation of divergent photon absorption in ultra-thin films of a topological insulator}
\author{Jing Wang}
\affiliation{Department of Physics, McCullough Building, Stanford University, Stanford, California 94305-4045, USA}
\author{Hideo Mabuchi}
\affiliation{Department of Applied Physics, Stanford University, Stanford, California 94305-4045, USA}
\author{Xiao-Liang Qi}
\affiliation{Department of Physics, McCullough Building, Stanford University, Stanford, California 94305-4045, USA}

\begin{abstract}
We perform linear and nonlinear photon absorption calculations in ultra-thin films of a topological insulator on a substrate.
Due to the unique band structure of the coupled topological surface states, novel features are observed for suitable photon frequencies,
including a divergent edge singularity in one-photon absorption process and a significant enhancement in two-photon absorption process.
The resonant frequencies can be controlled by tuning the energy difference and coupling of the top and bottom surface states. Such unique linear and nonlinear optical properties make ultra-thin films of topological insulators promising material building blocks for tunable high-efficiency
nanophotonic devices.
\end{abstract}

\date{\today}


\maketitle

\section{Introduction}
\label{introduction}

Time-reversal invariant topological insulators (TIs) are new states of quantum matter characterized by an insulating bulk state and gapless Dirac-type surface states.~\cite{qi2010,hasan2010,moore2010,qi2011} A range of compounds have been found to be three-dimensional (3D) TIs,~\cite{zhanghj2009,xia2009,chadov2010,lin2010,zhang2012} among which layered Bi$_2$Se$_3$ is demonstrated to be a prototype
3D TI with a large insulating bulk gap of about $0.3$~eV and metallic surface states with a single
Dirac cone.~\cite{zhanghj2009,xia2009} A thin layer of TI is expected to be a promising material for high-performance
optoelectronic devices such as photodetectors~\cite{zhangx2010} and transparent electrodes~\cite{penghl2012} due to its spin momentum locked massless Dirac surface state, which is topologically protected against time-reversal invariant perturbations.

Two-photon absorption (TPA) is a primary process of interest in various emergent photonics applications.~\cite{stryland1988,tutt1993,bravo2007,hayat2011,mabuchi2012}
For application purposes, a good TPA material must display large absorptive nonlinearities tuned within specific spectral regions.~\cite{wherrett1984,christodoulides2010}
To gain insight into the origin of large (degenerate) TPA coefficients $\beta$, we consider the expression for $\beta$ in second-order perturbation theory, which is proportional to the transition dipole moments and joint density of states (JDOS):
\begin{eqnarray}\label{TPA}
\beta(\omega) &\equiv& \frac{2\hbar\omega W_2}{\mathcal{I}^2}
\nonumber
\\
&=& \frac{2\hbar\omega}{\mathcal{I}^2}\frac{2\pi}{\hbar}\sum\limits_{\mathbf{k}}\left|\sum\limits_{i}\frac{\left\langle \psi_{c}\right|\mathcal{H}_1\left|\psi_i\right\rangle
\left\langle \psi_{i}\right|\mathcal{H}_1\left|\psi_v\right\rangle}{E_{i}(\mathbf{k}_{i})-E_{v}(\mathbf{k}_{v})-\hbar\omega}\right|^2
\nonumber
\\
&&\times\delta\left(E_{c}(\mathbf{k}_{c})-E_{v}(\mathbf{k}_{v})-2\hbar\omega\right),
\end{eqnarray}
where $\psi_{c}$, $\psi_{i}$ and $\psi_{v}$ are Bloch wavefunctions of the electrons in conduction, intermediate and valence bands, whose energies are $E_{c}(\mathbf{k}_{c})$, $E_{i}(\mathbf{k}_{i})$ and $E_{v}(\mathbf{k}_{v})$ and momenta are $\mathbf{k}_c$, $\mathbf{k}_i$ and $\mathbf{k}_v$. $\omega$ is the frequency of light. The delta function expresses energy conservation requirements for optical transition, and the summations over $i$ extend over all possible intermediate states. The $\mathbf{k}$ summation is over the entire first Brillouin zone, $\mathcal{H}_1$ is the electron-photon interaction Hamiltonian and $\mathcal{I}$ is the light irradiance. In general, one can get large $\beta$ when reaching the resonant condition ($E_{i}-E_{v}=2\hbar\omega$). Moreover, sharp peaks in the frequency dependence of the TPA coefficient should occur at critical points of the JDOS, such as Van Hove singularities. Two-dimensional (2D) systems may offer a novel avenue for creating useful TPA materials, unlike in 3D, Van Hove singularities in 2D may induce divergent JDOS.

\begin{figure}[b]
\begin{center}
\includegraphics[width=3.4in]{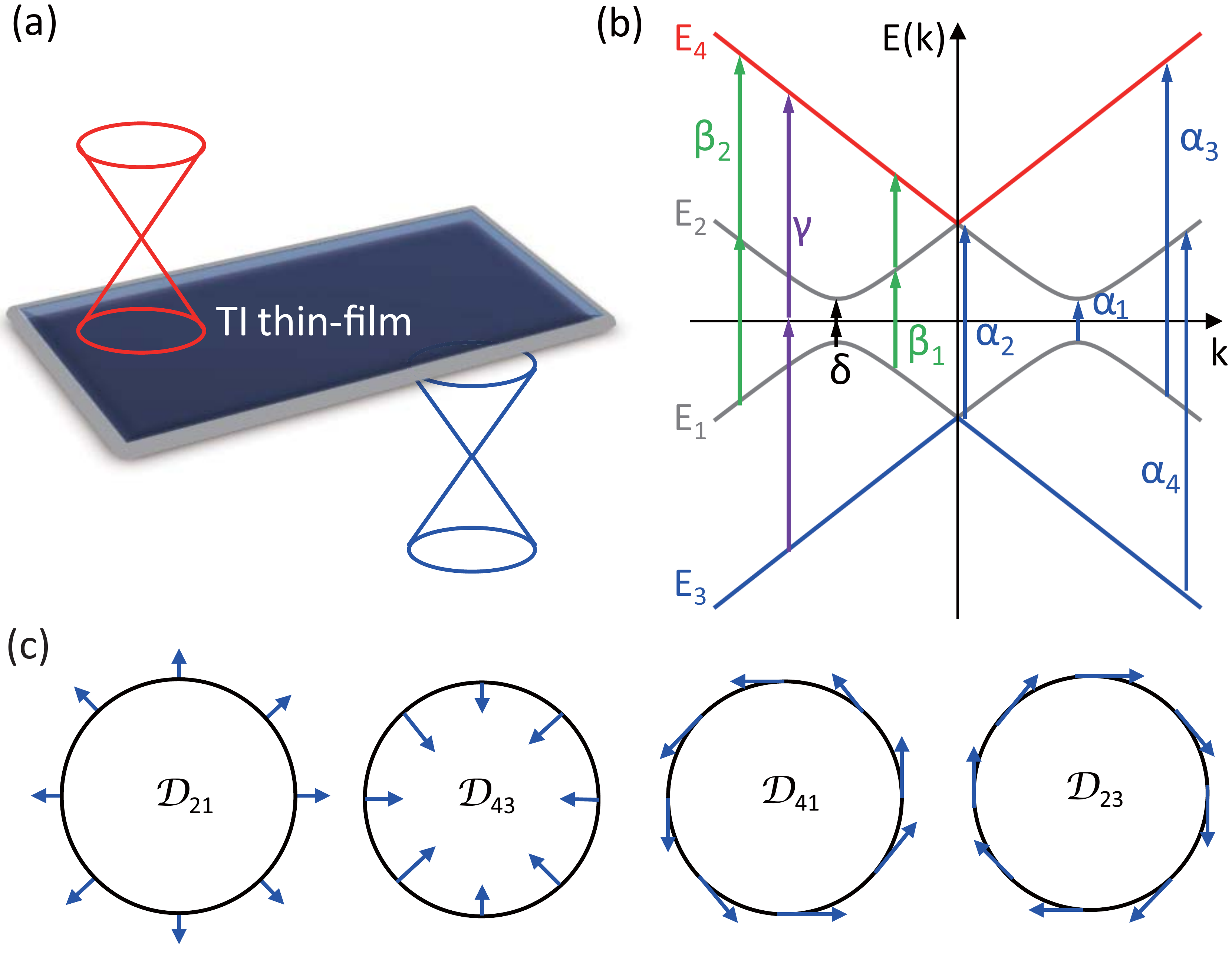}
\end{center}
\caption{(color online) (a) Coupling of Dirac cones on opposite surfaces of a thin-film TI. (b) Surface band structure with divergences in JDOS for one- ($\alpha_1$) and two-photon ($\delta$) transitions. The doping level of the system is in the gap. The bands and their labels are in the same color. The possible one-photon optical transitions are indicated by arrows with an index of $\alpha_1$, $\alpha_2$, $\alpha_3$, $\alpha_4$ and two-photon optical transitions by $\beta_1$, $\beta_2$, $\gamma$, $\delta$. (c) Optical selection rules for direct interband transition. $\mathcal{D}_{ji}$ denotes the optical transition from band $E_i$ to $E_j$. The ring denotes the constant energy contour for vertical transition, and arrow indicates the polarization of the optical field.}
\label{fig1}
\end{figure}

In this paper we show that thin films of a TI could provide a powerful setting, in which both linear and nonlinear optical processes of interest are greatly enhanced and are also highly tunable. A key feature of TIs is the existence of robust topological surface states, in which electrons propagate as massless relativistic fermions as shown in Fig.~\ref{fig1}(a). In an ultra-thin film (5~nm or thinner for Bi$_2$Se$_3$) of TIs the top and bottom surface states are coupled, giving rise to an energy gap.~\cite{linder2009,liu2010a,lu2010,zhangy2010} The coupling strength is controlled by the film thickness~\cite{liu2010a} and the energy difference between the two surface states can be controlled by substrate or electrical gating. Such tunability makes the TI thin film a unique 2D electron system. Due to the coupling of surface states, the conduction band minima and valence band maxima occur at the same nonzero wave-vector, leading to a divergent JDOS and thus a divergent one-photon absorption (OPA) at the gap frequency, illustrated by optical process $\alpha_1$ in Fig.~\ref{fig1}(b). Furthermore, by tuning the relative amplitude between the gap and the top-bottom surface energy difference, one can achieve a band structure in which a two-photon process [$\beta_1$ and $\beta_2$ in Fig.~\ref{fig1}(b)] is greatly enhanced due to the existence of an intermediate band at the resonance frequency and the almost divergent JDOS of the initial and final states [$\delta$ in Fig.~\ref{fig1}(b)]. With such properties, thin films of Bi$_2$Se$_3$ (and similar materials Bi$_2$Te$_3$ and Sb$_2$Te$_3$) are unique material building blocks for new nanophotonic devices.

The organization of this paper is as follows. After this
introductory section, Sec.~\ref{model} describes the model for the thin film of a TI and perturbative approach to
calculating the linear and nonlinear optical absorption. Section \ref{results} presents the results for OPA and TPA processes. Section \ref{discussion} presents the discussion. Section \ref{conclusion} concludes this paper.

\section{Model and Theory}
\label{model}

The low-energy physics of a TI thin film is characterized by two copies of the topological surface states on the top and bottom surfaces. In the simplest TIs such as the Bi$_2$Se$_3$ family, each surface state has a single Dirac cone. The surface state wavefunction is localized on the surface and decays exponentially away from the surface with a characteristic ``penetration depth" $\xi$. For the Bi$_2$Se$_3$ family of materials $\xi\sim 1$~nm. For ultra-thin films with thickness comparable with $\xi$, the overlap between the surface state wavefunctions from the two surfaces of the film become non-negligible and hybridization between them has to be taken into account. In a thin film TI on a substrate, the chemical potentials of the top and bottom surfaces are inequivalent and the Dirac points are generically at different energies. Considering the inter-surface coupling and the chemical potential difference one can write down the following low energy effective model of the thin film TI which matches well with experiment,~\cite{zhangy2010,shan2010}
\begin{equation}\label{Hamiltonian}
\mathcal{H}_0 = \hbar v\tau_z\otimes\left(\sigma_xk_y-\sigma_yk_x\right)+\frac{\Delta_{h}}{2}\tau_x\otimes1+\Delta_{ib}\tau_z\otimes1,
\end{equation}
where $v$ is the Dirac velocity, and $\sigma_{i}$ ($i=1,2,3$) and $\tau_j$ ($j=1,2,3$) are Pauli matrices acting on spin space and opposite surfaces, respectively. Time-reversal invariance follows from $\left[\Theta,\mathcal{H}_0\right]=0$, where $\Theta=1\otimes i\sigma_y\mathcal{K}$ and $\mathcal{K}$ is complex conjugation. $\Delta_{h}$ is the hybridization between the two surface states. $\Delta_{ib}$ is the inversion symmetry breaking, which can be substantially modified through electrical gating.~\cite{zhangy2010} Here for simplicity we neglect the higher-order terms in $k$, and we will discuss the effect of higher-order terms at the end of the paper. The surface band dispersion is
\begin{equation}\label{dispersion}
E(\mathbf{k}) =\mp\sqrt{\left(\hbar v\left|\mathbf{k}\right|\mp\Delta_{ib}\right)^2+\left(\Delta_{h}/2\right)^2},
\end{equation}
The energy gap is $E_{\text{edge}}=\Delta_{h}$ at the wavevector $\left|\mathbf{k}\right|=\Delta_{ib}/\hbar v$. In the following, we consider the doping level of the system is always in the gap.~\cite{grushin2012} $E_1$ and $E_3$ bands are occupied, while $E_2$ and $E_4$ bands are unoccupied. Thus the low energy optical absorption by the surface states can occur with photon energy ranging from $E_{\text{edge}}$ to gap of the bulk bands.

The direct electron-photon interaction is the dipole interband optical transitions, which is determined by minimal coupling, {\it i.e.}, by replacing $\mathbf{k}$ by $\mathbf{k}-e\mathbf{A}/\hbar c$ in the model Eq.~(\ref{Hamiltonian}), which leads to the interaction Hamiltonian
\begin{equation}
\mathcal{H}_1 =-\frac{e}{c}v\tau_z\otimes\left(\sigma_xA_y-\sigma_yA_x\right).
\end{equation}
Here $\mathbf{A}=A\mathbf{e}$ is the optical vector potential with amplitude $A$ and polarization $\mathbf{e}$. The amplitude $A$ is related to the light irradiance by $\mathcal{I}=\sqrt{\epsilon_{\omega}}\omega^2A^2/2\pi c$, and $\epsilon_{\omega}$ is the dielectric constant of the material. Here we have neglected the small wave vector of light. Taking into account the momentum conservation $\mathbf{k}$ for the initial $|\psi_{c}\rangle$ and final $|\psi_{v}\rangle$ states, only vertical excitation processes contribute to absorption.

The optical selection rules of the interband transition are obtained by the polarization operator
\begin{equation}
\mathcal{D}_{ji}(\mathbf{k}) \equiv \left\langle\psi_j(\mathbf{k})\right|\mathcal{H}_1\left|\psi_i(\mathbf{k})\right\rangle,
\end{equation}
here $\mathcal{D}_{ji}$ denotes the transition from state $|\psi_i\rangle$ to $|\psi_j\rangle$. The explicit form of the wavefuctions $|\psi_i\rangle$ ($i=1,2,3,4$) are
\begin{eqnarray}
|\psi_{1,2}(\mathbf{k})\rangle &=& \frac{1}{N_{1,2}}
\begin{bmatrix}
\frac{\left(\hbar v|\mathbf{k}|-\Delta_{ib}\right)\mp\sqrt{\left(\hbar v|\mathbf{k}|-\Delta_{ib}\right)^2+\Delta_h^2/4}}{\Delta_h/2}ie^{-i\theta_{\mathbf{k}}}\\
\frac{\left(\hbar v|\mathbf{k}|-\Delta_{ib}\right)\mp\sqrt{\left(\hbar v|\mathbf{k}|-\Delta_{ib}\right)^2+\Delta_h^2/4}}{\Delta_h/2}\\
ie^{-i\theta_{\mathbf{k}}}\\
1
\end{bmatrix}\nonumber
\\
|\psi_{3,4}(\mathbf{k})\rangle &=& \frac{1}{N_{3,4}}
\begin{bmatrix}
\frac{\left(\hbar v|\mathbf{k}|+\Delta_{ib}\right)\pm\sqrt{(\hbar v|\mathbf{k}|+\Delta_{ib})^2+\Delta_h^2/4}}{\Delta_h/2}ie^{-i\theta_{\mathbf{k}}}\\
-\frac{\left(\hbar v|\mathbf{k}|+\Delta_{ib}\right)\pm\sqrt{(\hbar v|\mathbf{k}|+\Delta_{ib})^2+\Delta_h^2/4}}{\Delta_h/2}\\
-ie^{-i\theta_{\mathbf{k}}}\\
1
\end{bmatrix}\nonumber
\end{eqnarray}
with $N_{1,2}$ and $N_{3,4}$ are normalization factor, and $\theta_{\mathbf{k}}$ is the azimuth angle between $\mathbf{k}$ and $k_x$-axis. Thus we have the transition for the surface bands $\mathcal{D}_{21}\propto\hat{\mathbf{k}}$, $\mathcal{D}_{43}\propto-\hat{\mathbf{k}}$, $\mathcal{D}_{41}\propto \hat{\mathbf{n}}\times\hat{\mathbf{k}}$ and $\mathcal{D}_{23}\propto-\hat{\mathbf{n}}\times\hat{\mathbf{k}}$, where $\hat{\mathbf{n}}$ is a unit vector normal to the momentum $\mathbf{k}$. Such selections rules is shown in Fig.~\ref{fig1}(c). We can see clearly that the optical selection rules in a thin film TI is very different from that in a conventional direct-gap semiconductor such as GaAs, where the spin wavefunction remains unchanged. This unique optical transition selection rule is due to the spin momentum locking of surface states.

The energy spectrum of the effective model (\ref{Hamiltonian}) is shown in Fig.~1(b). For finite chemical potential offset $\Delta_{ib}$, the coupling between the two surface states leads to avoided crossing of the energy dispersion at finite wavevectors, and the valence band maxima and conduction band minima coincide on a one-dimensional (1D) ring at $|{\bf k}|=\Delta_{ib}/\hbar v$ in momentum space. This feature is essential for optical properties of the system, since the density of states of both conduction and valence bands diverge at the same wavevectors, enabling a divergence in the probability of the optical transition process marked by $\alpha_1$ in Fig.~\ref{fig1}(b). Other important optical transitions beside $\alpha_1$ are also shown in Fig.~\ref{fig1}(b).

\section{Results}
\label{results}

\subsection{One-photon absorption}

\begin{figure}[b]
\begin{center}
\includegraphics[width=3.4in]{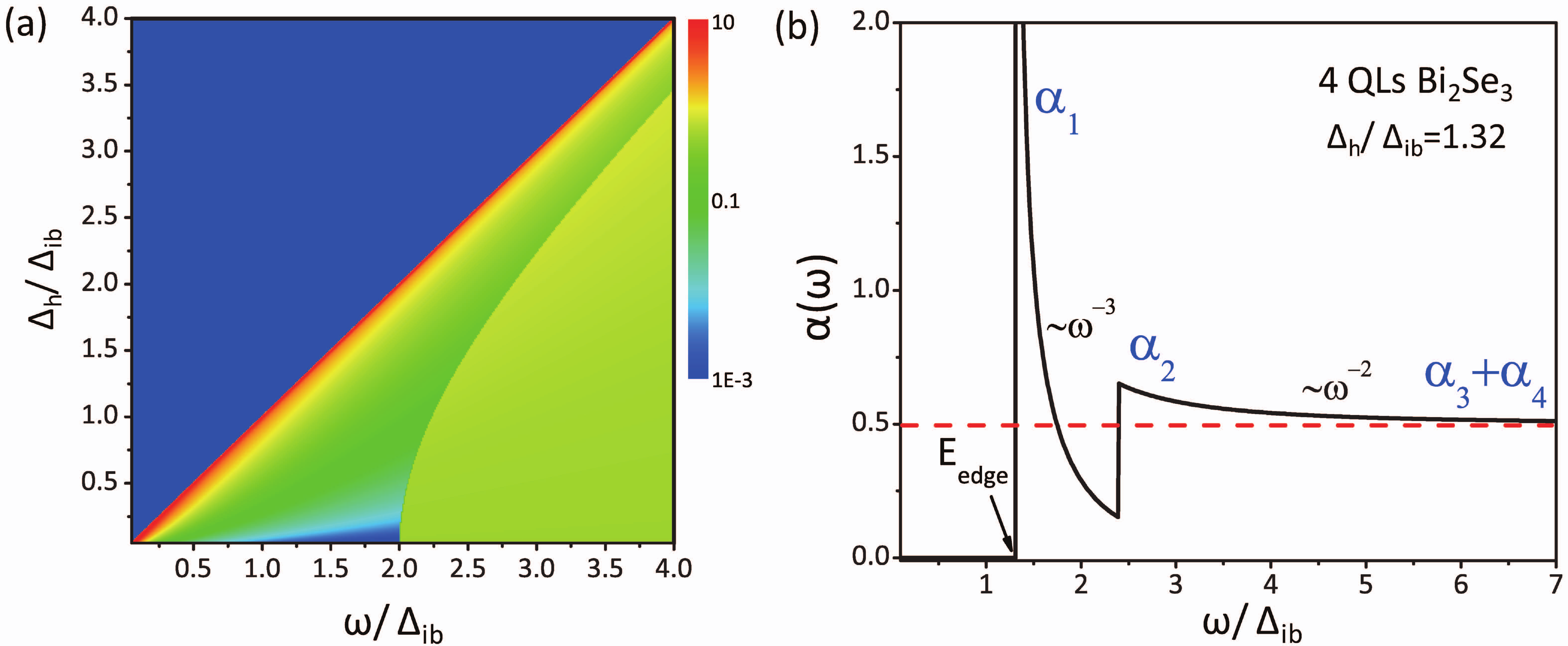}
\end{center}
\caption{ (color online) (a) Contour plot of OPA spectra for thin film TIs (logarithmic scale). $\Delta_{h}/\Delta_{ib}$ can be tuned by film thickness and electrical gating. $\alpha(\omega)$ is in units of $\pi\alpha/\sqrt{\epsilon_{\omega}}$. (b) Line cut for the 4-QLs Bi$_2$Se$_3$ thin film with $\Delta_{h}=70$~meV and $\Delta_{ib}=53$~meV. The edge of the interband transition $E_{\text{edge}}$ is indicated by an arrow; important features are labeled $\alpha_1-\alpha_4$.}
\label{fig2}
\end{figure}

The OPA coefficient is
\begin{equation}
\alpha(\omega) =\frac{\hbar\omega W_1}{\mathcal{I}},
\end{equation}
where $W_1$ is the transition probability rate for OPA per unit area
\begin{equation}\label{OPA}
W_1 =\frac{2\pi}{\hbar}\sum\limits_{\mathbf{k}}\sum\limits_{f\neq i}\left|\left\langle\psi_{f}\right|\mathcal{H}_1\left|\psi_{i}\right\rangle\right|^2
\delta\left(E_{fi}(\mathbf{k})-\hbar\omega\right),
\end{equation}
where $E_{fi}(\mathbf{k})\equiv E_{f}(\mathbf{k})-E_{i}(\mathbf{k})$. Fig.~\ref{fig2}(a) shows the OPA spectrum when the doping level is in the gap. Thus allowed optical transitions are $E_1\rightarrow E_2$, $E_3\rightarrow E_4$, $E_1\rightarrow E_4$, $E_3\rightarrow E_2$. These optical processes $\alpha_i$ ($i=1,2,3,4$) contribute to different features at different frequencies as marked in Fig.~\ref{fig2}(a). In the following we will discuss the contribution of these processes in more details.

\emph{1}. When the optical frequency $\hbar\omega<\Delta_{h}$, no real OPA will occur, for the energy conservation of the optical transition cannot be satisfied.

\emph{2}. As the frequency is larger at $\hbar\omega=\Delta_h$, the optical transition $\alpha_1$ can occur, and it will lead to the band edge singularity in OPA spectrum at the gap energy $E_{\text{edge}}$ as shown in Fig.~\ref{fig2}. This singularity is directly related to the JDOS divergence of the surface states in $E_1$ and $E_2$ bands. Explicitly, the summation
over the delta function $\delta(E_{21}(\mathbf{k})-\hbar\omega)$ in Eq.~(\ref{OPA}) can be converted into an integration over energy by
a JDOS,
\begin{equation}
\mathcal{N}(\omega) = \frac{1}{4\pi^2}\int \frac{dS_{\mathbf{k}}}{\left|\nabla_{\mathbf{k}}\left[E_2(\mathbf{k})-E_1(\mathbf{k})\right]\right|},
\end{equation}
where $S_{\mathbf{k}}$ is the constant energy surface defined by $E_{21}(\mathbf{k})=\hbar\omega$. When $\hbar\omega\leq\sqrt{4\Delta_{ib}^2+\Delta_{h}^2}$, the JDOS for $E_1$ and $E_2$ bands becomes
\begin{equation}
\mathcal{N}(\omega) = \frac{1}{2\pi}\frac{\hbar\omega}{\sqrt{\hbar^2\omega^2-\Delta_{h}^2}}.
\end{equation}
It becomes singular when $\hbar\omega=\Delta_h$ at finite wavevector. This square-root divergent Van
Hove singularity of the JDOS at the band edge is characteristic of one-dimensional behavior.~\cite{cappelluti2007prl}
For large photon frequencies, the $E_1\rightarrow E_2$ contribution to $\alpha(\omega)$ is proportional to $\omega^{-3}$.

\emph{3}. When the optical frequency is increased at $\hbar\omega=\sqrt{4\Delta_{ib}^2+\Delta_{h}^2}$, besides the $E_1\rightarrow E_2$ transition, other transitions $E_{1}\rightarrow E_4$, $E_{3}\rightarrow E_4$ and $E_{3}\rightarrow E_2$ start to occur at ${\bf k}=0$, as is labeled by $\alpha_2$ in Fig.~\ref{fig1}(b). These processes will lead to a step discontinuity in the OPA spectra. In particular, OPA from $E_{3}\rightarrow E_4$ is exactly zero at $\hbar\omega=\sqrt{4\Delta_{ib}^2+\Delta_{h}^2}$ and has $\omega^{-2}$ dependence when $\hbar\omega/\Delta_{ib}\gg1$.

\emph{4}. For frequency $\hbar\omega\gg\sqrt{4\Delta_{ib}^2+\Delta_{h}^2}$, the transitions $\alpha_3$ ($E_1\rightarrow E_4$) and $\alpha_4$ ($E_3\rightarrow E_2$) will occur far away from the avoided crossing wavevector $\Delta_{ib}/\hbar v$. In this limit the inter-surface coupling can be neglected, and the transition occurs within each surface. It has been studied in the graphene context~\cite{nair2008,wang2008,min2009} that such a transition in a 2D Dirac fermion system leads to a universal frequency-independent contribution $\pi\alpha/2\sqrt{\epsilon_{\omega}}$ to OPA with $\alpha\equiv e^2/\hbar c$ the fine-structure constant. This contribution dominates the absorption probability in the high frequency limit as shown in Fig.~\ref{fig2}(b).

\begin{figure*}[t]
\begin{center}
\includegraphics[width=\textwidth]{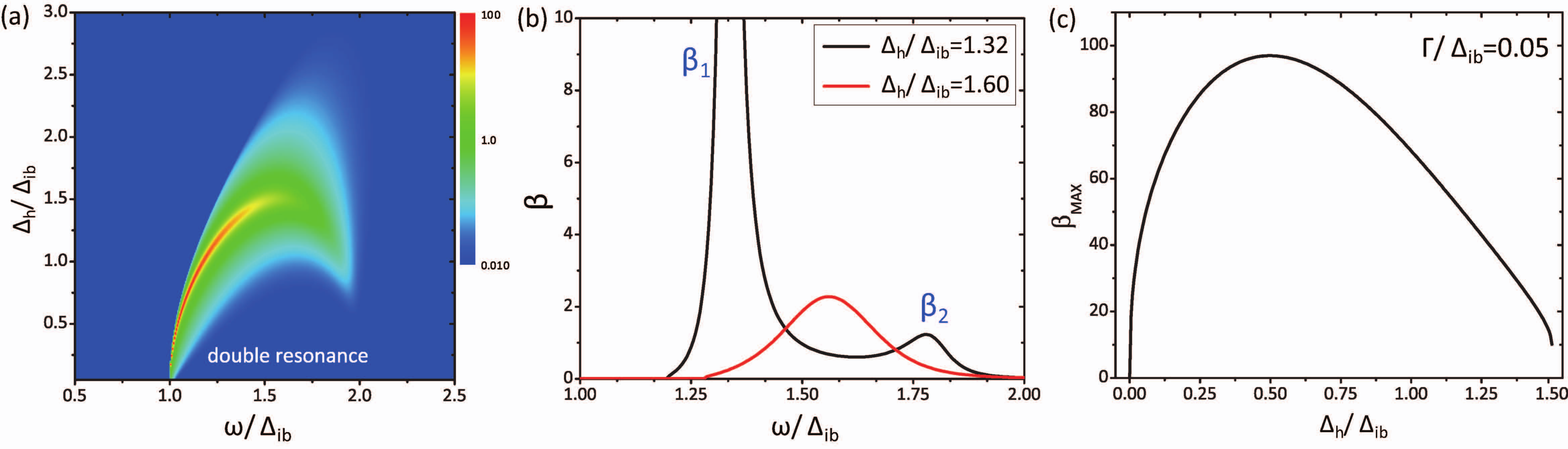}
\end{center}
\caption{ (color online) (a) Contour plot of TPA spectra of transitions from valence bands to $E_4$ for thin film TIs (logarithmic scale). $\Delta_{h}/\Delta_{ib}$ can be tuned by film thickness and electrical gating; $\Gamma/\Delta_{ib}=0.05$. $\beta$ is in units of $(4\pi^2\hbar/\epsilon_{\omega}\Delta_{ib}^4)(ve^2/c)^2$. (b) Line cut for $\Delta_{h}/\Delta_{ib}=1.32$ (4-QLs Bi$_2$Se$_3$) and $\Delta_{h}/\Delta_{ib}=1.6$. $\Delta_{h}/\Delta_{ib}=1.32$ has two resonances, labeled $\beta_1$ and $\beta_2$. (c) The maximum of $\beta$ vs.\ $\Delta_{h}/\Delta_{ib}$. It should be noticed that the resonance peak at $\hbar\omega=\Delta_h/2$ is not shown in this figure as it originates from the $E_1\rightarrow E_2$ transition (see Eq. (\ref{TPA12}) and the text).}
\label{fig3}
\end{figure*}

In short, as the frequency increases, the OPA spectrum of a TI thin film first has a square-root singularity at band edge $\hbar\omega=\Delta_h$, and then has a step discontinuity at $\hbar\omega=\sqrt{4\Delta_{ib}^2+\Delta_{h}^2}$, and approaches $\pi\alpha/2\sqrt{\epsilon_{\omega}}$ in the high frequency limit.

\subsection{Two-photon absorption}

For the surface state of a bulk TI, the TPA coefficient is obtained by including all possible intermediate states in the surface bands, which leads to
\begin{equation}
\beta_{\text{thick}} =\frac{2\pi^2}{\epsilon_{\omega}\omega^4\hbar^3}\left(\frac{ve^2}{c}\right)^2.
\end{equation}
There is no resonant feature or Van Hove singularity.

In a thin film new resonant features will appear due to the inter-surface coupling. There are
four allowed transitions, $E_1\rightarrow E_4\rightarrow E_2$ and $E_3\rightarrow E_4\rightarrow E_2$, $E_1\rightarrow E_2\rightarrow E_4$ ($\beta_1$, $\beta_2$) and $E_3\rightarrow E_2\rightarrow E_4$ ($\gamma$). We consider the case that the doping level is in the gap, so the latter two with nearly resonant condition dominate the optical process, as shown in Fig.~\ref{fig1}(b). Both transitions are included in the calculation of the TPA coefficient if $\hbar\omega\geq\sqrt{\Delta_{ib}^2+(\Delta_{h}/2)^2}$, as both of them satisfy energy conservation. In Fig.~\ref{fig3} we show numerical results for TPA coefficients and corresponding optical processes. When $\Delta_{h}/2<\hbar\omega<\sqrt{\Delta_{ib}^2+(\Delta_{h}/2)^2}$, the optical transitions from valence bands to conduction band $E_4$ is forbidden by energy conservation, so that $\beta=0$. As the photon frequency becomes larger, $\beta(\omega)$ has two resonance frequencies, corresponding to transitions $\beta_1$ and $\beta_2$, with the resonance condition $E_4-E_2=E_2-E_1=\hbar\omega$ indicated by Eq.~(\ref{TPA}). These features represent large tunable absorptive nonlinearities, making thin film TIs promising TPA materials for applications. The double resonance is at $k_{1,2}=(5\mu\pm\sqrt{9\Delta_{ib}^2-4\Delta_{h}^2})/4\hbar v$, and it disappears when $\Delta_{h}/\Delta_{ib}$ reaches a critical value $\Delta_{h}/\Delta_{ib}\geq1.5$ as shown in Fig.~\ref{fig3}(a). In practice, the energy $\hbar\omega$ in Eq.~(\ref{TPA}) needs to be replaced by $\hbar\omega+i\Gamma$ in order to take into account the effect of carrier damping. Here, $\Gamma$ is assumed to be a constant and inversely proportional to the dephasing time $\tau$. In our calculation we set $\Gamma/\Delta_{ib}=0.05$. Fig.~\ref{fig3}(b) shows the TPA spectrum for representative values of $\Delta_{h}/\Delta_{ib}$. $\beta(\omega)$ shows as double resonance feature for $\Delta_{h}/\Delta_{ib}=1.32$, which corresponds to the experimental values observed for 4 quintuple layers (QLs) Bi$_2$Se$_3$ film.~\cite{zhangy2010,shan2010} The strongest resonance feature occurs at $\beta_1$ since the optical transition matrix elements at $\beta_1$ are larger than that at $\beta_2$. In contrast, there is no strong resonance for $\Delta_{h}/\Delta_{ib}=1.60$. The process $\gamma$ does not satisfy the resonant condition for any photon frequency, so it has little contribution to the TPA coefficient. Fig.~\ref{fig3}(c) shows the maximum of $\beta$ (at $\beta_1$) versus the parameter $\Delta_{h}/\Delta_{ib}$. In particular, although the resonant condition is satisfied at $\omega=\Delta_{ib}$ when $\Delta_{h}=0$, the associated transition from $E_1\rightarrow E_2$ vanishes. The strongest $\beta$ occurs at $\Delta_{h}/\Delta_{ib}\approx0.5$.

The double resonance feature of the TPA coefficient from transition $E_1\rightarrow E_4$ is due to the Rashba-type splitting, compared to the single resonance in bilayer graphene.~\cite{yang2011} The Rashba splitting also gives rise to the divergent JDOS at the band edge, which has already been shown in the OPA coefficient. Obviously, there is no resonant intermediate states in the TPA process from $E_1\rightarrow E_2$, however, with the divergent JDOS and finite transition matrix elements at $\left|\mathbf{k}\right|=\Delta_{ib}$, the TPA contributed by the process $\delta$ around the gap is
\begin{equation}
E_1\rightarrow E_2:\ \ \beta(\omega) \propto \frac{1}{\omega^3\sqrt{\left(2\hbar\omega\right)^2-\Delta_{h}^2}}.\label{TPA12}
\end{equation}
It has a singular feature centered around $\hbar\omega=\Delta_{h}/2$ due to the Van Hove edge singularity. It shows $\omega^{-3}$ dependence in the resonant region $\hbar\omega\sim\Delta_{h}/2$ and $\omega^{-9}$ dependence in the off-resonant regions of $\hbar\omega\gg\Delta_h$, while TPA of gapless surface states in TIs has a $\omega^{-4}$ dependence for all photon frequency.

In short, at a given value of $\Delta_h$, the TPA spectrum of a TI thin film has a singularity at band edge $\hbar\omega=\Delta_h/2$. In addition, for small $\Delta_h$ there are two other resonance peaks appearing in the frequency range $\hbar\omega>\sqrt{\Delta_{ib}^2+(\Delta_{h}/2)^2}$ which merge into one peak and disappear at a certain $\Delta_h$.

\section{Discussion}
\label{discussion}

Taking into account of both OPA and TPA, the change in the intensity of the light
as it passes through the sample is given by
\begin{equation}
\Delta\mathcal{I}=-\alpha\mathcal{I}-\beta\mathcal{I}^2.
\end{equation}
The total absorption coefficient is given by $\alpha_{\text{total}}=\alpha+\beta\mathcal{I}$. The nonlinearity is characterized by the ratio $\beta\mathcal{I}/\alpha$ which depends on the intensity $\mathcal{I}$. Since for the TI film we have shown that $\beta$ has resonance features occurring at frequencies different from that of $\alpha$, the ratio $\beta/\alpha$ can be greatly enhanced, enabling the realization of strong nonlinearity at low intensity of light.
For the 4-QL Bi$_2$Se$_3$, the parameters are estimated by $\Delta_{ib}=53$~meV.~\cite{sobota2012,giraud2011} In our calculation we have taken $\tau\sim2.3$~ps, which leads to $\Gamma=2.9$~meV. For the resonant frequency of $\beta$ $1.25<\hbar\omega/\Delta_{ib}<1.8$, the condition $\beta\mathcal{I}/\alpha\sim1$ can be satisfied for an (optical) electric field strength of $10^5$~V/m. Such field strengths potentially could be realized at low incident optical powers using photonic resonators with high ratio of quality-factor to mode-volume.~\cite{yee2009}

The interaction of surface states with phonons and impurities will cause electron relaxation. In fact, the dephasing time of the topological surface states observed in experiments.may be longer than the $\tau$ we take. From the femtosecond time- and angle-resolved photoemission spectroscopy,~\cite{sobota2012} a long-lived population of a metallic Dirac surface state ($>10$~ps) in Bi$_2$Se$_3$ has been found. Such long dephasing time is directly related to the spin texture of surface states. The electron-phonon interaction in TI is also weak because the small Fermi surface limits the number of phonon modes coupled with electrons. According to Ref. \onlinecite{giraud2011}, the broadening is $\Gamma<1$~meV at $T<50$~K.

In general, the electron hole pair at two surfaces under external bias will have Coulomb interaction. In the mean field approximation, the interaction here would prefer the exciton condensate when the chemical potential is smaller than $\Delta_h$.~\cite{seradjeh2009} This is exactly the case considered here that the doping level is always in the gap. (It should be pointed out that there are many interesting effects when the chemical potential is in the conduction band, which need further investigation.~\cite{stauber2007}) Such exciton condensate will enhance the hybridization of the two surface states, and thus the resonance frequency of the system.~\cite{zhangy2010}

There are higher order terms such as hexagonal warping term proportional to $k^3$ in the surface state dispersion relation.~\cite{fu2009} With such terms the gap due to avoid crossing of surface states is no longer uniform around the crossing wavevectors, and the JDOS becomes finite. However, the warping parameter is very small in Bi$_2$Se$_3$ when the crossing energy is lower than 0.22~eV (defined respect to the Dirac point), so that the JDOS enhancement given by our low energy effective theory remains valid.~\cite{wang2011}

It worths mentioning that the gap of TI thin films is always less than the bulk gap 0.3~eV, which is in the THz frequency range. However, Raman processes may lead to transition between $E_1$ and $E_{2,4}$ in the optical frequency range where the intermediate states are high-energy bulk states,~\cite{jenkins2010,hsieh2011b} such Raman process should be greatly enhanced due to the divergent JDOS and could be more conveniently accessed in experiments and applications.

\section{Conclusion}
\label{conclusion}

In conclusion, the TI thin film has interesting linear and non-linear optical properties.
The unique band structure of the coupled topological surface states gives rise to a divergent edge singularity in one-photon absorption process and a significant enhancement in two-photon absorption process. The tunable one-photon and two-photon absorption in this system may find applications in spintronics, such as optical generation of spin current and charge current.~\cite{zhao2006} Such unique linear and nonlinear optical properties make ultra-thin films of TIs promising material building blocks for tunable high-efficiency nanophotonic devices.

\acknowledgments
We are grateful to Y. Cui, H. Y. Hwang, R. B. Liu and S. C. Zhang for insightful discussions. This work is supported by the Defense Advanced Research
Projects Agency Microsystems Technology Office, MesoDynamic Architecture Program (MESO) through contract Nos.
N66001-11-1-4105 (J. W. and X. L. Q.) and N66001-11-1-4106 (H. M.), and by the US Department of Energy, Office of Basic Energy Sciences, Division of Materials Sciences and Engineering, under contract No. DE-AC02-76SF00515.

\end{document}